\documentclass[twoside,12pt]{article}

\usepackage{apacite}
\usepackage{obs_study_style}
\usepackage{amsmath}
\usepackage[normalem]{ulem}
\newcommand{\rpm}{\raisebox{.2ex}{$\scriptstyle\pm$}}

\usepackage[bottom]{footmisc}
\usepackage{bbm, xcolor, color}
\usepackage[T1]{fontenc}
\usepackage{enumitem}
\usepackage[mathscr]{euscript}
 \let\mathscr\relax
\usepackage[scr]{rsfso}

\usepackage{multicol}
\usepackage{setspace}



\usepackage{etoolbox}
\newcounter{bibcount}
\makeatletter
\patchcmd{\@lbibitem}{\item[}{\item[\hfil\stepcounter{bibcount}{\thebibcount.}}{}{}
\renewcommand\NAT@bibsetup%
   [1]{\setlength{\leftmargin}{\bibhang}\setlength{\itemindent}{-\parindent}%
       \setlength{\itemsep}{\bibsep}\setlength{\parsep}{\z@}}
\makeatother

\usepackage{xcolor}     
\usepackage{lipsum}     
\usepackage{ntheorem}   
\usepackage{mdframed}   

\theoremstyle{break}
\theoremheaderfont{\bfseries}
\newmdtheoremenv[%
linecolor=lightgray,leftmargin=30,%
rightmargin=30,
backgroundcolor=gray!40,%
innertopmargin=4pt,%
ntheorem]{myprop}{}[section]

\heading{}{}{}{}{}{H. Campbell and P. Gustafson}
\ShortHeadings{Conditional Equivalence Testing: an alternative remedy to publication bias}{Campbell and Gustafson}
\firstpageno{1}

\begin{document}

\title{Conditional Equivalence Testing: an alternative remedy for publication bias}

\author{\name Harlan Campbell  \email harlan.campbell@stat.ubc.ca \\
	\AND
         \name Paul Gustafson \email gustaf@stat.ubc.ca \\
         \\
     \addr Department of Statistics, University of British Columbia, Vancouver, Canada
       }

\maketitle

{\centering October, 2017 \par}

\vspace{1cm}

\begin{abstract}
We introduce a publication policy that incorporates ``conditional equivalence testing'' (CET), a two-stage testing scheme in which standard NHST is followed conditionally by testing for equivalence.  The idea of CET is carefully considered as it has the potential to address recent concerns about reproducibility and the limited publication of null results.  In this paper we detail the implementation of CET, investigate similarities with a Bayesian testing scheme, and outline the basis for how a scientific journal could proceed to reduce publication bias while remaining relevant.

\end{abstract}

\begin{keywords}
  Null Hypothesis Significance Testing, reliability, equivalence testing, non-inferiority testing, Bayesian testing, statistical power, confidence intervals
\end{keywords}

\section{Introduction}

Poor reliability, within many scientific fields, is a major concern for researchers, scientific journals and the public at large.  In a highly cited essay, \citet{ioannidis2005most} uses Bayes theorem to claim that more than half of published  research findings are false.  While not all agree with the extent of this conclusion, the argument raises concerns about the trustworthiness of science, amplified by a disturbing prevalence of scientific misconduct, \citet{fanelli2009many}.  That the reliability of a result may be substantially lower than its $p$-value suggests has been ``underappreciated'' \citep{goodman2007assessing}, to say the least.  

To address this problem, some journal editors have taken radical measures (e.g. \citet{TrafimowMarks2015}), while the reputation of researchers and the credibility of the science they generate are tarnished \citep{wasserstein2016asa}.  The foundational premise of null hypothesis significance testing (NHST) is now in question\footnote{ To be clear, NHST has long been controversial \citep{nickerson2000null, harlow2016if} but this controversy has recently been renewed.} \citep{lash2017harm, hofmann2016null, szucs2016null, cumming2014new}, and the usefulness of the $p$-value vigorously debated \citep{lew2013p, chavalarias2016evolution, gelman2013commentary}. 

The limited publication of null results is certainly one of the most substantial factors contributing to low reliability.  Whether due to a reluctance of journals to publish null results or to a reluctance of investigators to submit their null research \citep{dickersin1992factors, reysen2006publication}, the consequence is severe publication bias; see \citet{franco2014publication} and \citet{doshi2013restoring}.    Despite repeated warnings, publication bias persists and, to a certain degree, this is understandable.  Accepting a null result can be difficult, owing to the well-known fact that ``absence of evidence is not evidence of absence'' \citep{hartung1983absence, altman1995statistics}.  As \citet{greenwald1975consequences} writes: ``it is inadvisable to place confidence in results that support a null hypothesis because there are too many ways (including incompetence of the researcher), other than the null hypothesis being true, for obtaining a null result.''   Indeed, this is the foremost critique of NHST, that it cannot provide evidence in favour of the null hypothesis.  The commonly held belief that for non-significant result to show high ``retrospective power'' \citep{zumbo1998note} implies support in favour of the null, is problematic; see \citet{hoenig2001abuse}.  In fact, a larger $p$-value (e.g. $p$-value $>$ 0.05), combined with high power often occurs even in situations when the data support the alternative hypothesis more than the null, \citet{greenland2012nonsignificance}.

In order to address publication bias, it is often suggested \citep{walster1970proposal, sterling1995publication, dwan2008systematic, sune2013positive} that publication decisions should be made without regards to the statistical significance of results, i.e. ``result-blind peer review'' \citep{greve2013result}.  In fact, a growing number of psychology and neuroscience journals are adopting pre-registration \citep{nosek2017preregistration} including ``Registered Reports'' (RR) \citep{chambers2015registered}, a publication policy in which authors ``pre-register their hypotheses and planned analyses before the collection of data'' \citep{chambers2014instead}.  If the rationale and methods are sound, the RR journal agrees (before any data are collected) to publish the study regardless of the eventual data and outcome obtained.  Among many potential pitfalls with result-blind peer review \citep{findley2016can}, a legitimate and substantial concern is that, if a journal were to adopt such a policy, it might quickly become a ``dumping ground'' \citep{greve2013result} for null and ambiguous findings that do little to contribute to the advancement of science.  To address these concerns, RR journals require that, for any manuscript to be accepted, authors must provide a-priori (before any data are collected) sample size calculations that show statistical power of at least 90\% (in some cases 80\%).   This is a reasonable remedy to a difficult problem.  Still, it is problematic for two reasons.  

First, it is acknowledged that this policy will disadvantage researchers who work with ``expensive techniques or who have limited resources'' \citep{chambers2014instead}.  While not ideal, small studies can  provide definitive value and potential for learning; see \citet{sackett1993can}.  For this reason, some go as far as arguing against any requirement for a-priori sample size calculations (i.e. needing to show sufficient power as a requisite for publication).   For example, \citet{bacchetti2002peer} writes: ``If a study finds important information by blind luck instead of good planning, I still want to know the results''; see also \citet{aycaguer2013explicacion}.
While unfortunate, the loss of potentially valuable ``blind luck'' results and small sample studies \citep{matthews1995small} appears to be a necessary price to pay for keeping a ``result-blind peer review'' journal relevant.  Is this too high a price?  Based on simulations, \citet{borm2009publication} conclude that the negative impact of publication bias does not warrant the exclusion of studies with low power.


Second, a-priori power calculations are often flawed, due to the unfortunate ``sample size samba'' \citep{schulz2005sample}:  the practice of retrofitting the anticipated effect size in order to obtain a desirable sample size.   Even under ideal circumstances, a-priori power estimation is often ``wildly optimistic'' \citep{bland2009tyranny}  and heavily biased due to the ``illusion of power'' \citep{vasishth2017illusion}.  This ``illusion'' occurs when the estimated effect size is based on a literature filled with overestimates (to be expected in many fields due, somewhat ironically,  to publication bias).  \citet{djulbegovic2011optimism} conduct a retrospective analysis of phase III randomized controlled trials (RCTs) and conclude that optimism bias significantly contributes to inconclusive results; see also \citet{chalmers2006implications}.  What's more, oftentimes due to unanticipated difficulties with enrolment, the actual sample size achieved is substantially lower than the target set out a-priori, \citet{chan2008discrepancies}.  In these situations, RR requires that either the study is rejected/withdrawn for publication or a certain leeway is given under special circumstances (Chambers (2017), personal communication).  Neither option is ideal.  Given these difficulties with a-priori power calculations, it remains to be seen to what extent the 90\% power requirement will reduce the number of underpowered publications that could lead a journal to be a dreaded ``dumping ground''.

An alternative proposal to address publication bias and the related issues surrounding low reliability is for researchers to adopt Bayesian testing schemes; e.g. \citet{dienes2017four}, \citet{ kruschke2017bayesian} and \citet{wagenmakers2007practical}.  It has been suggested that with Bayesian methods, publication bias will be mitigated ``because the evidence can be measured to be just as strong either way'' \citep{dienes2016bayes}.   Bayesian methods may also provide for a better understanding of the strength of evidence \citep{etz2016bayesian}.  However, researchers in many fields remain uncomfortable with the need to define (subjective) priors and are concerned that Bayesian methods may increase ``researcher degrees of freedom'' \citep{simmons2011false}.  Furthermore, it is acknowledged that sample sizes will typically need to be larger than with equivalent frequentist testing in situations when there is little prior information incorporated, \citet{zhang2011bayesian}.  Nevertheless, a number of RR journals allow for a Bayesian option.  At registration (before any data are collected), rather than committing to a specific sample size, researchers commit to attaining a certain Bayes Factor (BF).  For example, the journals \emph{Comprehensive Results in Social Psychology} (CRSP) and \emph{NFS Journal} (the official journal of the Society of Nutrition and Food Science) require that one pledges to collect data until the BF is more than 3 (or less than 0.33), \citet{jonassubmission}. The journals \emph{BMC Biology} and the \emph{Journal of Cognition} require, as a requisite for publication, a BF of at least 6 (or less than 1/6)\footnote{See editorial policies at: https://bmcbiol.biomedcentral.com/about/registered-reports, and https://www.journalofcognition.org/about/registered-reports/}.

In this paper, we propose an alternate option made possible by adapting NHST to conditionally incorporate equivalence testing.  While equivalence testing is by no means a novel idea, previous attempts to introduce equivalence testing have ``largely failed'' \citep{lakens2017equivalence}.  Our proposal to systematically incorporate equivalence testing into a two-stage testing procedure has not been extensively pursued (one exception may be \citet{hauck1986proposal}) and there has not been any discussion of how such a testing procedure could facilitate publication decisions for peer-review journals.  One reason for this is a poor understanding of the \emph{conditional} equivalence testing strategy whereby testing traditional non-equivalence (or superiority) is followed conditionally, by testing equivalence (or non-inferiority).  In fact, whether or not such a two-stage approach is beneficial has been somewhat controversial.  As the sample-size is typically determined based only on the primary test, the power of the secondary equivalence (non-inferiority) test is not controlled, thereby potentially increasing the false discovery rate, \citet{ng2003issues}. \citet{koyama2005decision} investigate and conclude that, in most situations, such concern is unwarranted.   In Section 2, in order to \mbox{further} the understanding of \emph{conditional} equivalence testing, we provide a brief overview including how to carry out power calculations and how to (not necessarily prior to the study) establish appropriate equivalence  margins. 


One reason conditional equivalence testing (CET) is an appealing approach is that it shares many of the properties that make Bayesian testing schemes so attractive.  As such, the publication policy we put forward is somewhat similar to the RR ``Bayesian option''.  With conditional equivalence testing, evidence can be measured in favour of both the alternative and the null (at least in a pragmatic sense), and as such is ``compatible with a Bayesian point of view'' \citep{ocana2008equivalence}.  In {Section 3}, we conduct a simple simulation study to demonstrate how one will often arrive at the same conclusion whether using CET or a Bayesian testing scheme.  In {Section 4}, we outline how a publication policy could be framed around CET to encourage the publication of null results and make recommendations for reporting and implementation.  Finally, {Section 5} concludes with suggestions for future research.

\section{Conditional Equivalence Testing Overview}

Standard equivalence testing is essentially NHST with the hypotheses reversed.  For example, for a two-sample study of means, the equivalence testing null hypothesis would be a difference in means, and the alternative hypothesis would be equal (within a given margin) means.  \emph{Conditional} equivalence testing (CET) is the practice of standard NHST followed conditionally (if one fails to reject the null) by equivalence testing.  
CET is not an altogether new way of testing.  Rather it is the usage of established testing methods in a way that permits better interpretation of results.  In this regard, it is similar to other proposals such as the ``three-way testing'' scheme proposed by \citet{goeman2010three}, and \citet{zhao2016considering}'s proposal for incorporating both statistical and clinical significance into one's testing.  To illustrate, what follows is a brief outline of CET for a two-sample test of equal means (assuming equal variance).

Let $x_{i1}$, for $i=1,...,n_{1}$ and $x_{i2}$, for $i=1,...,n_{2}$ be independent random samples from two normally distributed populations of interest with  $\mu_{1}$, the true mean of population 1; $\mu_{2}$, the true mean of population 2; and $\sigma^{2}$, the true common population variance.  Let $n=n_{1}+n_{2}$ and define sample means and sample variances as follows:
$\bar{x}_{g}=\sum_{i=1}^{n_{g}}x_{gi}$,  and \mbox{$s^{2}_g=\sum_{i=1}^{n_{g}}(x_{gi}-\bar{x}_{g})^2/(n_g-1)$}, for $g =1,2.$  Also, let \mbox{$s_{p} =\sqrt{((n_{1}-1)s^{2}_1 + (n_{2}-1)s^{2}_2)/(n_{1}+n_{2}-2)}$}.  The true difference in population means, $\mu_{d}=\mu_{1} - \mu_{2}$, under the standard null hypothesis, $H_{0}$, is equal to zero.  Under the standard alternative, $H_{1}$, we have that $\mu_{d} \ne 0$.

The term equivalence is not used in the strict sense that $\mu_{1}=\mu_{2}$.  Instead, equivalence in this context refers to the notion that the two means are ``close enough'', i.e. their difference is within the \emph{equivalence margin}, $\delta =[-\Delta, \Delta]$, chosen to define a range of values considered equivalent (i.e. the ``zone of indifference'').  In equivalence (and non-inferiority) trials, the $\Delta$ is ideally chosen to be the ``minimum clinically meaningful difference'' \citep{kaul2006good, greene2008noninferiority}.   

Let $F_{T_{df}}()$ be the cumulative distribution function (cdf) of the $t$ distribution with $df$ degrees of freedom and define the following critical $t$ values: $ t^{*}_{\alpha_{1}/2} = F^{-1}_{T_{n-2}}(1-0.5\alpha_{1}) $ (i.e. the upper 100$\cdot{\frac{\alpha_{1}}{2}}$-th percentile of the t-distribution with $n-2$ degrees of freedom) and  $ t^{*}_{\alpha_{2}} = F^{-1}_{T_{n-2}}(1-\alpha_{2})$,  (i.e. the upper 100$\cdot{\alpha_{2}}$-th percentile of the $t$-distribution with $n-2$ degrees of freedom).  As such, $\alpha_{1}$ is the maximum allowable type I error (e.g. $\alpha_{1}$=0.05) and $\alpha_{2}$ is the maximum allowable ``type $E$'' error (erroneously concluding equivalence), possibly equal to $\alpha_{1}$.  If a type $E$ error is deemed less costly than a type I error, $\alpha_{2}$ may be set higher (e.g. $\alpha_{2}$=0.10).  CET is the following conditional procedure consisting of five steps:



\begin{myprop}
\begin{small}
\begin{quote}
\textbf{Step 1- A two-sided, two-sample $t$-test for a difference of means.} \\
\indent  \indent Calculate the $t$-statistic, $T = (\bar{x}_{1}-\bar{x}_{2})/(s_{p}\sqrt{1/n_1 + 1/n_2})$ and associated $p$-value, $p_{1}=2\cdot F_{T_{n-2}}(-|T|)$.

\textbf{Step 2-}
\indent \indent If $ |T| > t^{*}_{\alpha_{1}/2}$, then declare a \emph{positive result}.  There is evidence of a statistically significant difference, $p$-value = $p_{1}$. \\
\indent \indent Otherwise, if $ |T| \le t^{*}_{\alpha_{1}/2}$, proceed to Step 3.

\textbf{Step 3- Two one-sided tests (TOST) for equivalence of means.} \\
\indent \indent Calculate two $t$-statistics: $T_1 =  (\bar{x}_{1}-\bar{x}_{2}+\Delta)/(s_{p}\sqrt{1/n_1 + 1/n_2})$, and  $T_2 =  (\bar{x}_{1}-\bar{x}_{2}-\Delta)/(s_{p}\sqrt{1/n_1 + 1/n_2})$.  Calculate an associated $p$-value,  $p_2= max(F_{T_{n-2}}(-T_{1}),F_{T_{n-2}}(T_{2}))$.  Note: $p_{2}$ is a \emph{marginal} $p$-value, in that it is not calculated under the assumption that $p_{1}>\alpha_{1}$.  
 \\
\textbf{Step 4- }
\indent \indent If $T_1 >  t^{*}_{\alpha_{2}}$ and  $T_2 < (-t^{*}_{\alpha_{2}})$, declare a \emph{negative result}. There is evidence of a statistically significant equivalence ($\delta =[-\Delta, \Delta]$), \mbox{$p$-value = $p_{2}$. } \\
\indent \indent Otherwise, proceed to Step 5.

\textbf{Step 5-} Declare an \emph{inconclusive result}.  There is insufficient evidence to support any conclusion.
\end{quote}
\end{small}
\end{myprop}

For ease of explanation, let us define $p_{CET}= p_{1}$ if the result is positive, and $p_{CET}= 1- p_{2}$ if the result is negative or inconclusive.  Thus, a small value of $p_{CET}$ suggests evidence in favour of a positive result, whereas a large value of $p_{CET}$  suggests evidence in favour of a negative result.  As is noted above, it is important to acknowledge that, despite the above procedure being dubbed ``conditional'', $p_{2}$ is a \emph{marginal} $p$-value, i.e. it is not calculated under the assumption that $p_{1}>\alpha_{1}$.  The interpretation of $p_{2}$ would be the same regardless of whether it was obtained following Steps 1 and 2 or was obtained ``on its own'' via standard equivalence testing.
 
 Standard NHST involves the same first and second steps and ends with an alternative Step 3 which states that if  $p_{1}$>$\alpha_{1}$, one declares an inconclusive result (`there is insufficient evidence to reject the null.').  Similar to two-sided CET, one-sided CET is straightforward, making use of non-inferiority testing in Step 3.  Note that one-sided CET testing, like all one-sided testing, is vulnerable to potential post-hoc abuse, i.e. the direction of the test could be based on the data \citep{freedman2008analysis}.  

There is a large literature on Step 3's TOST and non-inferiority testing, see \citet{walker2011understanding} and \citet{meyners2012equivalence} for overviews that cover the basics as well as more subtle issues.  There are also many proposed alternatives to TOST for equivalence testing, what are known as ``the emperor's new tests'' \citep{perlman1999emperor}.   These alternative are offered as marginally more powerful options, yet are more complex and are not widely used.  As such, we will not go into any further detail and refer those interested to \citet{meyners2012equivalence}.

CET is a procedure applicable to any type of outcome.  Let $\theta$ be the parameter of interest in a more general testing setting.  CET can be described in terms of calculating confidence intervals around $\hat{\theta}$, the statistic of interest.  For example, $\theta$ may be defined as the the difference in sample proportions, the hazard ratio, the risk ratio, or the slope of a linear regression model, etc.  For details on equivalence testing in more general scenarios (i.e. non-continuously distributed outcomes,  one-sample and two-sample tests), see \citet{chen2000tests} (equivalence testing for two proportions),  \citet{da2009methods} (binary and survival outcomes), \citet{wiens2000design} (three treatment arms), \citet{wellek2010testing} (binary outcomes, count outcomes and more), \citet{dixon2005statistical} (linear trends), \citet{dannenberg1994extension} (data with unequal variances), and \citet{hauschke1990distribution} (nonparametric tests).  Note that in these other cases, the equivalence margin, $\delta$, may not necessarily be centred at zero, but will be a symmetric interval around  $\theta_{0}$, the value of $\theta$ under the standard null hypothesis.  In general, let $\delta=[\delta_L, \delta_U]$, with $\Delta$ equal to half the length of the interval.    Consider, more generally, CET as the following procedure: 

\begin{myprop}
\begin{small}
\begin{quote}
\textbf{Step 1-} Calculate a $(1-\alpha_{1})\%$ Confidence Interval for $\theta$. 

\textbf{Step 2-}  If this C.I. excludes $\theta_{0}$, then declare a \emph{positive result}.\\
\indent \indent Otherwise, if $\theta_{0}$ is within the C.I., proceed to Step 3.

\textbf{Step 3-} Calculate a $(1-2\alpha_{2})\%$ Confidence Interval for $\theta$. 

\textbf{Step 4-} If this C.I. is entirely within $\delta$, declare a \emph{negative result}.  \\
\indent \indent Otherwise, proceed to Step 5.

\textbf{Step 5-} Declare an \emph{inconclusive result}.  \\
\indent \indent There is insufficient evidence to support any conclusion.
\end{quote}
\end{small}
\end{myprop}


\begin{figure}[h!]
\begin{centering}
\fbox{\includegraphics[width=12cm]{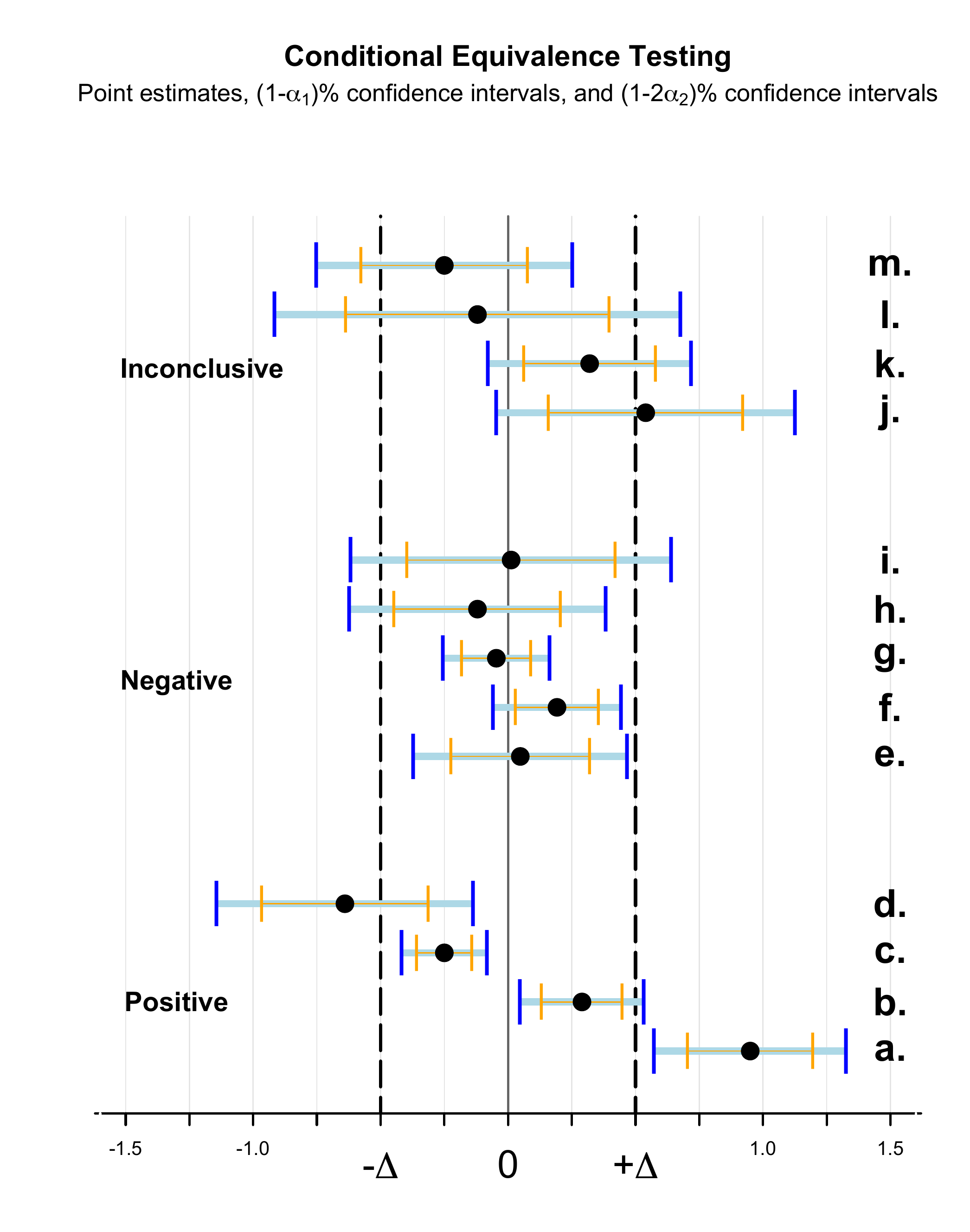}}
\caption{\footnotesize{Point estimates and confidence intervals of thirteen possible results from two-sample testing of normally distributed data (as in box 2.1) are presented alongside their corresponding conclusions.  Here $\theta_{0}=0$ and $2\alpha_{2}>\alpha_{1}$.  Black points indicate point estimates; blue lines (wider intervals) represent (1-$\alpha_{1}$)\% confidence intervals; and orange lines (shorter intervals) represent (1-2$\alpha_{2}$)\% confidence intervals.}}
\label{fig:CET1}
\end{centering}
\end{figure}

Figure \ref{fig:CET1} illustrates the three conclusions with their associated confidence intervals.  In this figure, we once again consider two-sample testing of normally distributed data as described for box 2.1., with $\alpha_{1}=0.05$ and $\alpha_{2}=0.10$.  Situation ``a'', in which the $(1-\alpha_{1})\%$ C.I. is entirely outside the equivalence margin is what \citet{guyatt1995basic} calls a ``definitive-positive'' result.  The lower confidence limit of the parameter is not only larger than zero (the null value, $\theta_{0}$), implying a ``positive''  result, but also is above the $\Delta$ threshold.  Situations ``b'' and ``c'', in which the  $(1-\alpha_{1})\%$ C.I. excludes zero and the $(1-2\alpha_{2})\%$ C.I. is within $[-\Delta,\Delta]$ are considered ``positive'' results but require some additional interpretation.   One could  describe the effect in these cases as ``significant yet not meaningful'' or conclude that there is evidence of a significant effect, yet the effect is likely ``not minimally important''.  A  ``positive'' result such as ``d'', with a wider confidence interval, represents a significant, albeit imprecisely estimated, effect.  One could conclude that additional studies, with larger sample sizes, are required to determine if the effect is of a meaningful magnitude.

Evidently, in some cases, the three categories are not sufficient on their own for adequate interpretation.  For example, without any additional information the ``positive'' vs. ``negative'' distinction between cases ``b'' and ``f'' is misleading.  These two results appear very similar: in both cases a substantial (i.e. greater than $\Delta$-sized) effect can be ruled out.  Indeed, positive result ``b'' appears much more like the negative result ``f'' than positive result ``a''.  Additional language and careful attention to the estimated effect size is required for correct interpretation.  While case ``k'' has a similar point estimate to ``b'' and ``f'', the wider C.I. ($\Delta$ is within $(1-2\alpha_{2})\%$ C.I.) means that one cannot rule out the possibility of a meaningful effect, and so it is rightly categorized as ``inconclusive''.

It may be argued that CET is simply a recalibration of the standard confidence interval, like converting Fahrenheit to Celsius.  This is valid commentary and in response, it should be noted that our suggestion to adopt CET (in the place of standard confidence intervals) is not unlike that of \citet{gardner1986confidence} who suggest that confidence intervals should replace $p$-values; see also \citet{cumming2008replication} and \citet{reichardt2016confidence}.   One advantage of CET over confidence intervals is that it may improve the interpretation of null results, see \citet{parkhurst2001statistical} and \citet{hauck1986proposal}.    By clearly distinguishing between what is a \emph{negative} versus an \emph{inconclusive} result, CET serves to simplify the long ``series of searching questions''  necessary to evaluate a ``failed outcome'' \citep{pocock2016primary}.   However, as can be seen with the examples of Figure 1, the use of CET should not rule the complementary use of confidence intervals.  Indeed, the best interpretation of a result will be when using both tools together. 

\begin{figure}[h!]
\begin{centering}
\fbox{\includegraphics[width=15cm]{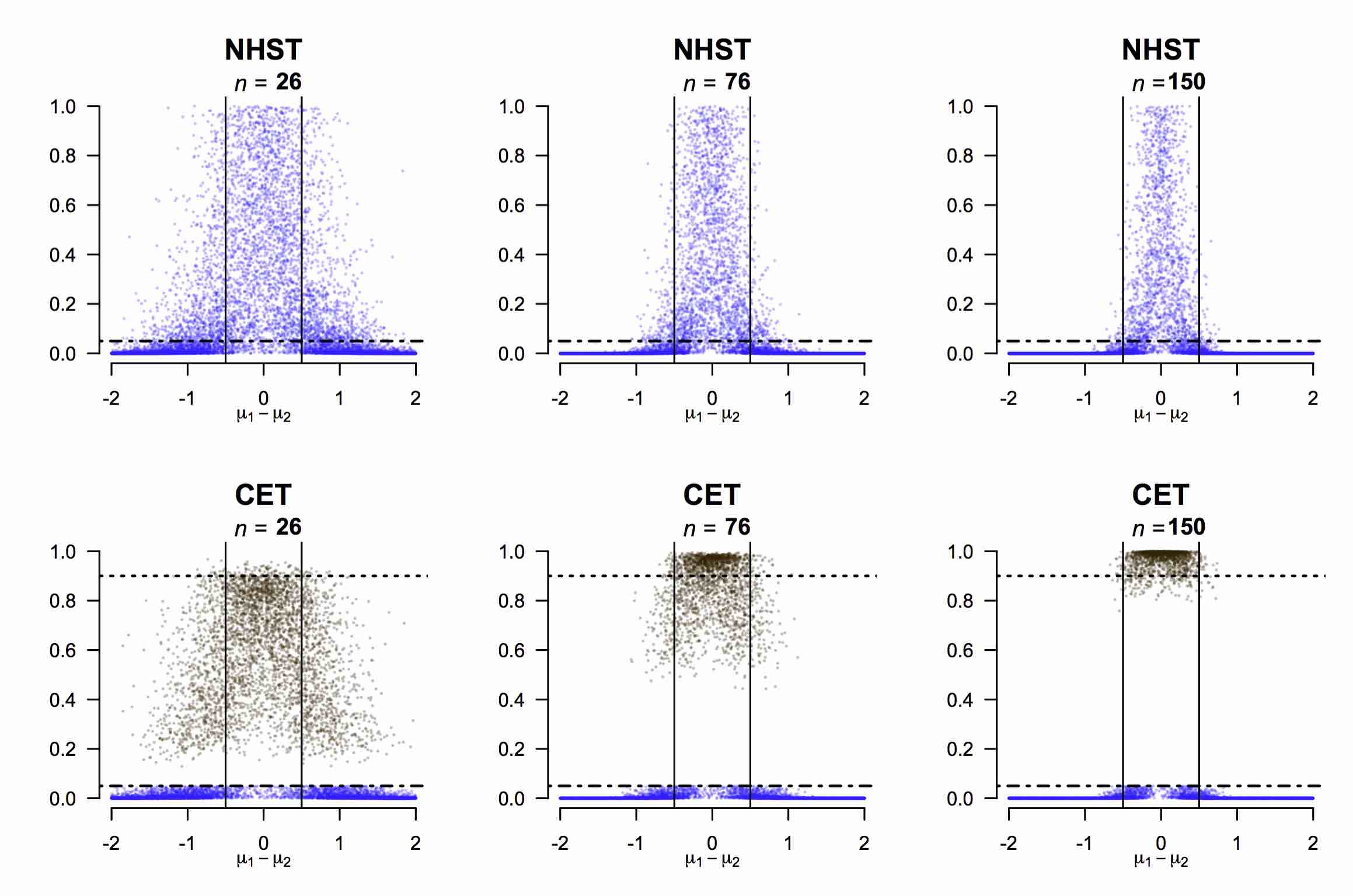}}
\caption{\footnotesize{ Distribution of two-tailed $p$-values from NHST and $p_{CET}$ values from CET, with varying $n$ (total sample size) and $\delta=[-0.5, 0.5]$. Data are the results from 10,000 Monte Carlo simulations of two-sample normally distributed data with equal variance, $\sigma^{2}=1$.  The true difference in means, $\mu_{d}=\mu_{1}-\mu_{2}$, is drawn from a uniform distribution between -2 and 2.  Black points $(=1-p_2)$  indicate evidence in favour of equivalence (``negative result''), whereas blue points ($=p_1$) indicate evidence in favour of non-equivalence (``positive result'').   An ``inconclusive result'' would occur for all black points falling in between the two dashed horizontal lines at $\alpha_{1}=0.05$ and $\alpha_{2}=0.10$, (i.e for $p_2>\alpha_2$).  The format of this plot is based on Figure 1, ``Distribution of two-tailed P-values from Student's $t$-test'', from \citet{lew2013p}.}}
\label{fig:Lew_ALT2}
\end{centering}
\end{figure}

As \citet{dienes2017four} clearly explain, with standard NHST, one is unable to make the ``three-way distinction'' between the positive, inconclusive and negative (``evidence for $H_{1}$'', ``no evidence to speak of'', and ``evidence for $H_{0}$'').  Figure~\ref{fig:Lew_ALT2} shows the distribution of standard two-tailed $p$-values under NHST and corresponding $p_{CET}$ values under CET for three different sample sizes.  These are the result of two-sample testing of normally distributed data as described for box 2.1, with $n_{1}=n_{2}$. 

Under the standard null ($\mu_{d}=0$) with NHST, one is just as likely to obtain a large $p$-value as one is to obtain a small $p$-value.  Under the standard null with CET, a small $p_2$ value (a large $p_{CET}$ value) will indicate evidence in favour of equivalence given sufficient data.  In Figure~\ref{fig:Lew_ALT2}, note that the blue points that fall below the $\alpha_{1}=0.05$ threshold in the upper panels (NHST) remain unchanged in the lower panels (CET).  However, blue points that fall above the $\alpha_{1}=0.05$ threshold in the upper panels (NHST) are no longer present in the lower panels (CET).  They have been replaced by black points ($=p_{CET}= 1-p_{2}$) which, if near the top, suggest evidence in favour of equivalence.  This treatment of the larger $p$-values is more conducive to the interpretation of null results bringing to mind the thoughts of \citet{amrhein2017earth} who write ``[a]s long as we treat our larger $p$-values as unwanted children, they will continue disappearing in our file drawers, causing publication bias, which has been identified as the possibly most prevalent threat to reliability and replicability.''

\subsection{Defining the equivalence margin}

As with standard equivalence and non-inferiority testing, defining the equivalence margin will be one of the ``most difficult issues'' for CET \citep{hung2005regulatory}.  If  the margin is too large, then a claim of equivalence is meaningless. If the margin is too small, then the probability of declaring equivalence, will be substantially reduced; see \citet{wiens2002choosing}.   As stated earlier, the margin is ideally chosen as a boundary to exclude `minimum clinically meaningful differences' \citep{kaul2006good, greene2008noninferiority}.  However, ``clinically meaningful'' effects are difficult to define, and there is generally no clear consensus among stakeholders, \citet{keefe2013defining}.  Furthermore, previously agreed-upon meaningful differences may be difficult to ascertain as they are rarely specified in protocols and published results \citep{djulbegovic2011optimism}. 

In some fields, there are some generally accepted norms.  For example, in bioavailability studies, equivalence is routinely defined (and listed by regulatory authorities) as a difference of less than 20\%.  In oncology trials, a small effect size has been defined as odds ratio or hazard ratio of 1.3 or less, \citet{bedard2007statistical}.  In ecology, a proposed equivalence region for trends in population size (the log-linear population regression slope) is $\delta = [-0.0346, 0.0346]$, \citet{dixon2005statistical}.  

In cases when a specified equivalence margin may not be as clear-cut, less conventional options have been put forth.  \citet{hauck1986proposal} propose the concept of using an ``equivalence curve'' that illustrates results for a range of possibilities.  \citet{meyners2007least} proposes to use the least equivalent allowable difference (LEAD), the largest possible value of $\Delta$ for which one can claim equivalence.  The public would then be left to draw their own conclusions from the data and whether they believe the LEAD-$\Delta$ is a reasonable standard for equivalence.  This would no doubt lead to a discussion about which effect sizes are too small to be worthwhile for a given treatment, and advance researchers towards a specific standard.    

One important choice a researcher must make in defining the equivalence margin is whether the margin should be defined on a raw or standardized scale. \citet{lakens2017equivalence} discusses the pros and cons of each option.  For example, if there is no rationale for any standard margin in our two-sample normal case, taking equivalence to be a difference within half the estimated standard deviation, i.e. defining $\Delta=qs_{p}$, with pre-specified $q$=0.5 seems reasonable.  Note, that the probability of obtaining a negative result may be zero (or negligible) for certain combinations of values of $\alpha_1$, $\alpha_{2}$, $n_1$, $n_2$ and $q$.  For example, with $q=0.5$, $\alpha_1=0.05$, and $\alpha_{2}=0.10$, $Pr(negative)=0$ for all $n\le26$ (with $n_1=n_2$).  As such, $q$ must be chosen with additional practical considerations (see additional details in 2.2).  For binary and time-to-event outcomes, there is an even greater number of different ways one can define the margin (e.g. in terms of relative risk vs. odds ratio).  For discussions on this, see \citet{ng2008noninferiority}, \citet{da2009methods}, \citet{tsou2007mixed} and \citet{barker2001equivalence}.  


It is important to note that, while it may be ideal to specify the margin prior to collecting data, setting the margin afterwards will not lead to any type I error inflation (i.e. one will not erroneously reject $H_{0}:  \theta=\theta_{0}$ with probability greater than $\alpha_{1}$).  For the same reason that a 95\% C.I. is just as valid as a 85\% C.I., but must be interpreted differently, CET is valid regardless of whether the margin is specified (on a raw, or a standardized scale by defining $q$) before or after data are obtained.  However, the conclusion made will clearly depend upon the margin chosen, and for any non-positive result, it is always be possible to choose a specific margin (or specific value for $q$) which leads to a negative conclusion (see Figure \ref{fig:CET1}).  Since the choice of margin is often a difficult one in the best of circumstances, a retrospective choice is not ideal as there will be ample room for bias in one's choice, regardless of how well intentioned one may be.  For this reason, for equivalence and non-inferiority RCTs, it is generally expected that margins are to be pre-specified \citep{piaggio2006reporting}.

\subsection{Operating characteristics and sample size calculations}

What follows is a brief overview of how to calculate the probabilities of obtaining each of the three conclusions (positive, negative, and inconclusive) as listed in our CET procedure above for two-sample testing of normal data with equal variance (box 2.1).  More in-depth related work includes \citet{shieh2016exact} who establishes exact power calculations for TOST for equivalence of normally distributed data and  \citet{da2009methods} who review power and sample size calculations for equivalence testing of binary and time-to-event outcomes.

\begin{figure}[h!]
\begin{centering}
\fbox{\includegraphics[width=13cm]{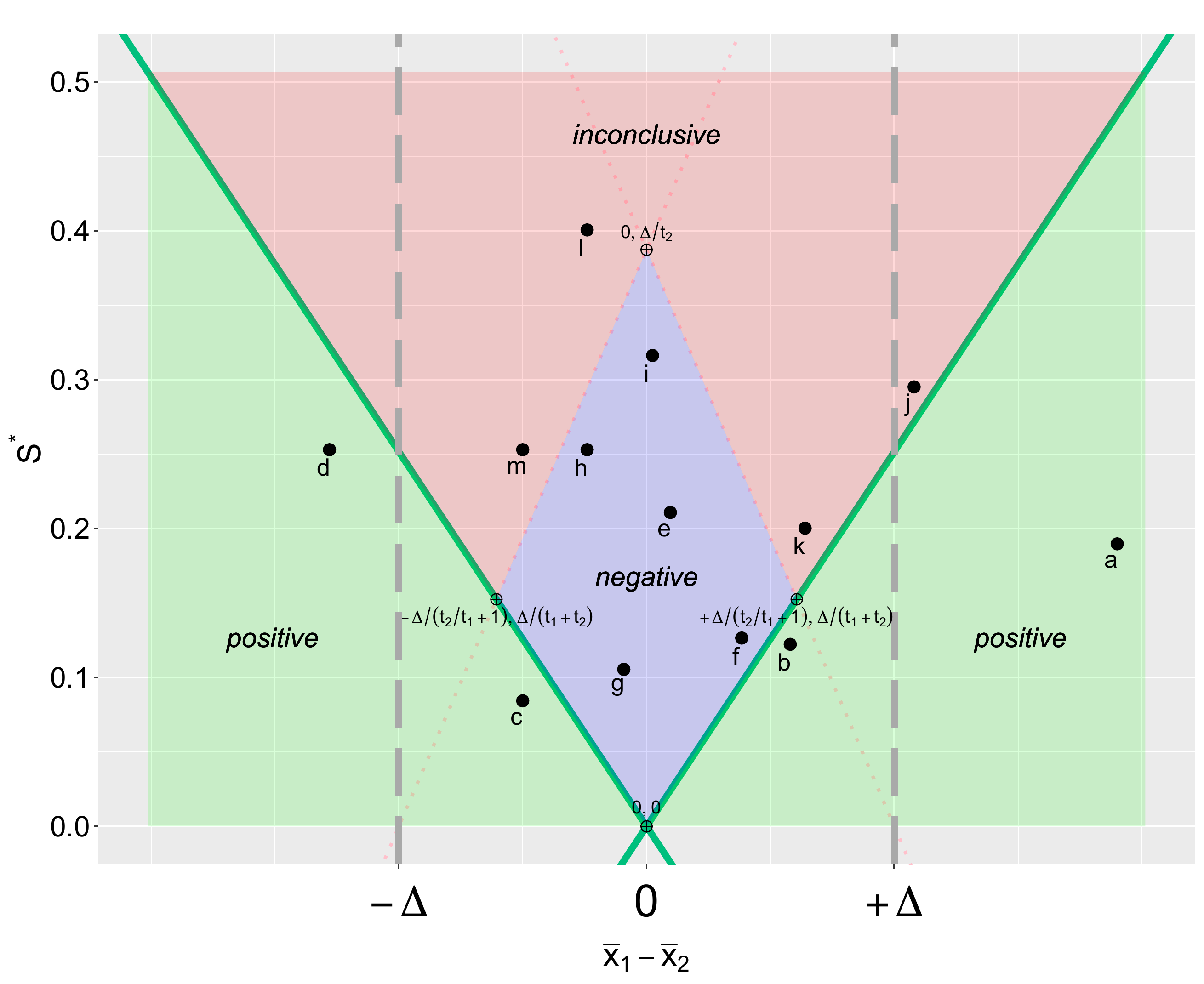}}
\caption{\footnotesize{Let $n=90$, $\Delta=0.5$, $t_{1}=t^{*}_{\alpha_{1}/2}$, $t_{2}=t^{*}_{\alpha_{2}}$.  The values of $\hat{\mu}_{d}=\bar{x}_{1}-\bar{x}_{2}$ (on the  $x$-axis) and $s^{*}$ (on the $y$-axis) are obtained from the data.  The three conclusions, positive, negative, inconclusive correspond to the three areas shaded in green, blue and red respectively. The black lettered points correspond to the scenarios of Figure \ref{fig:CET1}.}  }
\label{fig:diamond}
\end{centering}
\end{figure}

As before, let the true population mean difference be $\mu_{d} = {\mu}_{1}-{\mu}_{2}$ and the true population variance equal ${\sigma}^{2}$.  Let $\sigma^{*}=\sigma\sqrt{(1/n_{1} +1/n_{2})}$ and $s^{*}=s_{p}\sqrt{(1/n_{1} +1/n_{2})}$.  Figure \ref{fig:diamond} illustrates how each of the three conclusions can be reached based on the values of $s^{*}$ and $\hat{\mu}_{d}=\bar{x}_{1}-\bar{x}_{2}$ obtained from the data.  For this plot, $n=90$ (with $n_{1}=n_{2}$), $\Delta=0.5$, $\alpha_{1}=0.05$ and $\alpha_{2}=0.10$.  The black lettered points correspond to the scenarios of Figure \ref{fig:CET1}.  The interior diamond, $\Diamond$, (with corners located at [0,0], $[-\Delta/( t^{*}_{\alpha_{2}}/t^{*}_{\alpha_{1}/2} + 1 ) , \Delta/( t^{*}_{\alpha_{1}/2}+t^{*}_{\alpha_{2}} )], [0,\Delta/t^{*}_{\alpha_{2}}]  , [\Delta/( t^{*}_{\alpha_{2}}/t^{*}_{\alpha_{1}/2} + 1 ) , \Delta/( t_{1}+t_{2} )]$) covers the values for which a negative conclusion is obtained.  A positive conclusion corresponds to when $|\bar{x}_{1} - \bar{x}_{2}|$ is large and $s^{*}$ is relatively small. Note how $\Delta$ and ${\sigma}$ impact the conclusion. If the equivalence margin is sufficiently wide, one will have $Pr(inconclusive)$ approach zero as $\sigma$ approaches   zero.  Indeed, the ratio of ${\Delta}/{\sigma}$ determines, to a large extent, the probability of a negative result.  If the equivalence margin is sufficiently narrow and the variance relatively large (i.e. ${\Delta}/{\sigma}$ is very small), one will have $Pr(negative)\approx0$.

Let us assume for simplicity that $n_{1}=n_{2}$.  Then, the sampling distributions of the sample mean and sample variance are well established:\mbox{ $\hat{\mu}_{d} \sim N(\mu_{d}, \sigma^{*2}) $} and $\frac{(n -2)s_{p}^{2}}{\sigma^{2}} \sim \chi^{2}_{n-2}$.  Therefore, given fixed values for $\mu$ and $\sigma^{2}$, we can calculate the probability of obtaining a positive result, $Pr(positive)$.   In Figure \ref{fig:diamond}, $Pr(positive)$ equals the probability of $\hat{\mu}_{d}$ and $s^{*}$ falling into either the left  or right ``positive'' corners and is calculated (as in a usual power calculation for NHST):

\begin{equation}  
 Pr(positive; \mu_{d}, \sigma) = \Big(1-F_{n-2, \frac{\mu_{d}}{\sigma^{*}}} \Big( t^{*}_{\alpha_{1}/2} \Big) \Big) + F_{n-2, \frac{\mu_{d}}{\sigma^{*}}} \Big(-t^{*}_{\alpha_{1}/2} \Big) 
\end{equation}

\noindent where $F_{df, ncp}(x)$ is the cdf of the non-central $t$ distribution with $df$ degrees of freedom and non-centrality parameter $ncp$.

One can calculate the probability of obtaining a negative result, $Pr(negative)$, as the probability of $\hat{\mu}_{d}$ and $s^{*}$ falling into the ``negative'' diamond, $\Diamond$.  Since $\hat{\mu}_{d}$ and $s^{*}$ are independent statistics, we can write their joint density as the product of a normal probability density function, $f_{N}()$, and a chi-squared probability density function, $f_{\chi^{2}}()$. However, the resulting double integral will remain difficult to evaluate algebraically over the boundary, $\Diamond$.  Therefore, the probability is best approximated numerically, for example, by Monte Carlo integration, as follows:

\begin{eqnarray}  
 Pr(negative; \mu_{d}, \sigma)  &=& \int\int_{\Diamond} f_{N}(u; \mu_{d},  \sigma^{*2})f_{\chi^{2}}(v; n-2) du dv    \nonumber \\
  &=&  \int_{\Diamond} \Big({\Phi(h_{2}(v); \mu_{d},  \sigma^{*})-\Phi(h_{1}(v); \mu_{d},  \sigma^{*})\Big)f_{\chi^{2}}(v; n-2)dv}  \nonumber \\  
  &\approx&  \sum_{j=1}^{M}\Big({\Phi(h_{2}(c_{j}); \mu_{d},  \sigma^{*})-\Phi(h_{1}(c_{j}); \mu_{d},  \sigma^{*})\Big) /M \nonumber}
\end{eqnarray}

\noindent where $\Phi()$ is the normal cdf and Monte Carlo draws from a chi-squared distribution provide $c_{j} = \sqrt{\sigma^{2} q^{[j]}/(n-2)} $, with $q^{[j]}\sim \chi^{2}_{n-2}$ for $j = 1,..., M$.  The left and right-hand boundaries of the diamond-shaped ``negative region'', are defined by $h_{1}(c_{j})=min(0,max(+c_{j}t_2-\Delta, -c_{j}t_1))$ and $h_{2}(c_{j}) = max(0,min(-c_{j}t_{2}+\Delta, +c_{j}t_{1}))$. 

Defining the boundary with $h_{1}()$ and $h_{2}()$ allows for three distinct cases as seen in Figure \ref{fig:diamond}: 

\noindent (1) $s^{*}>\Delta/t^{*}_{\alpha_{2}}$, in which case $h_{1}(s^{*})=h_{2}(s^{*})=0$;

\noindent (2) $\Delta/(t^{*}_{\alpha_{1}/2}+t^{*}_{\alpha_{2}})<s^{*}<\Delta/t^{*}_{\alpha_{2}}$, in which case $h_{1}(s^{*}) = -\Delta + t^{*}_{\alpha_{2}}s^{*}$ and $h_{2}(s^{*})=\Delta - t^{*}_{\alpha_{2}}s^{*}$; and

\noindent (3) $s^{*}<\Delta/(t^{*}_{\alpha_{1}/2}+t^{*}_{\alpha_{2}})$, in which case $h_{1}(s^{*}) = -t^{*}_{\alpha_{1}/2}s^{*}$ and $h_{2}(s^{*}) = +t^{*}_{\alpha_{1}/2}s^{*}$. \\

\begin{figure}[h!]
\begin{centering}
\fbox{\includegraphics[width=13cm]{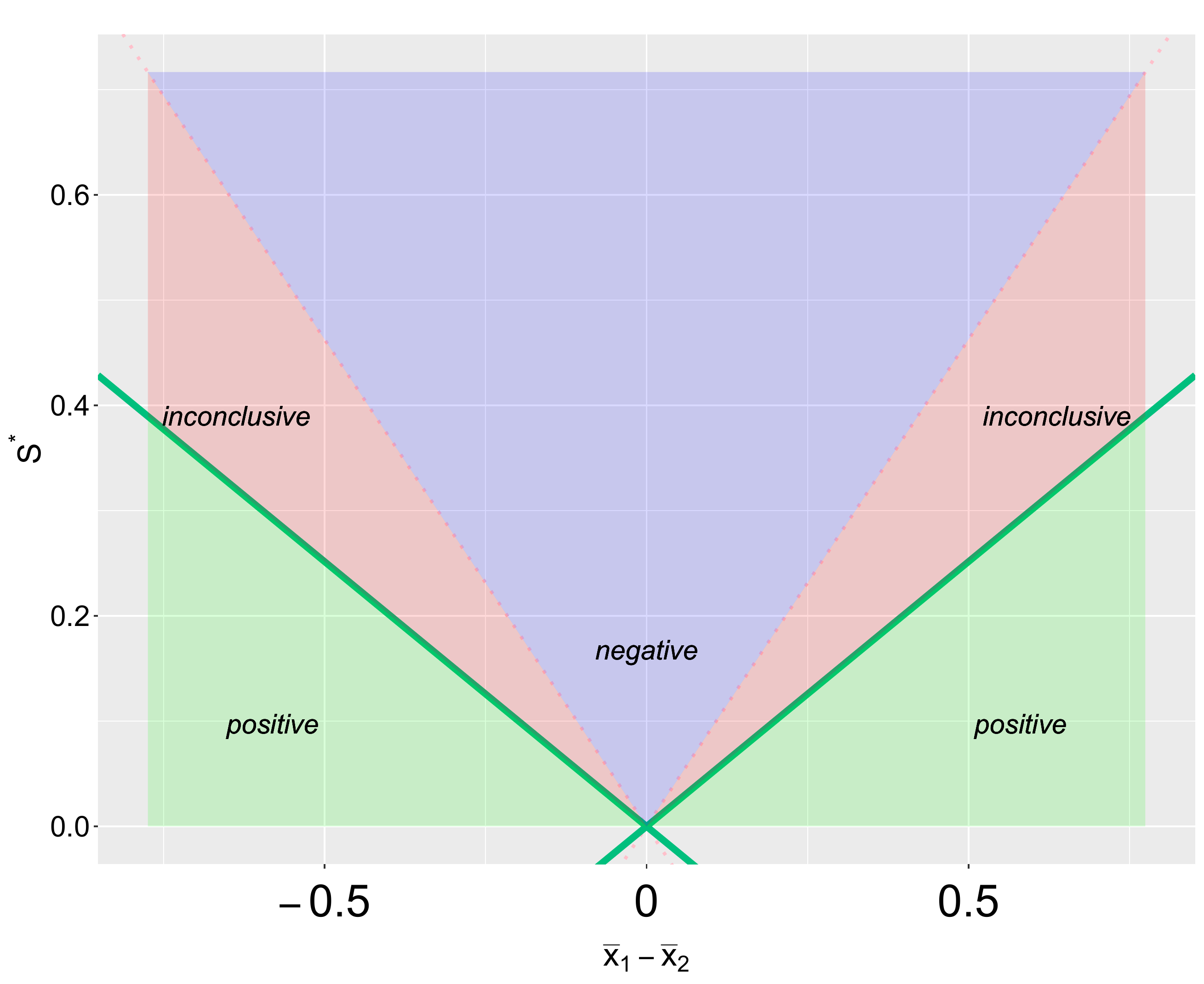}}
\caption{\footnotesize{Let $n=90$, $\Delta=0.5s_{p}$, $\alpha_{1}=0.05$ and $\alpha_{2}=0.10$. The three conclusions, positive, negative, inconclusive correspond to the three areas shaded in green, blue and red respectively.}}
\label{fig:diamond2}
\end{centering}
\end{figure}

When the equivalence boundaries are defined as a function of $s_{p}$ (e.g. $\Delta=qs_{p}$), the calculations are somewhat different;  Figure \ref{fig:diamond2} illustrates.  In particular,  a negative conclusion requires: $ qs^{*}/\sqrt{1/n_1+1/n_2} > |\hat{\mu}_{d} \rpm s^{*}t^{*}_{\alpha_{2}} |$ (i.e. the $(1-2\alpha_{2})$\% C.I. is entirely within $[-\Delta,\Delta]$). As such, for a given sample size, it will only be possible to obtain a negative result if  $q > t^{*}_{\alpha_{2}}\sqrt{1/n_1+1/n_2}$.  Likewise, an inconclusive result will only be possible if $(q/\sqrt{1/n_1+1/n_2} ) < (t^{*}_{\alpha_{1}/2} + t^{*}_{\alpha_{2}} )$.

In order to determine an appropriate sample size for CET, one must replace $\mu_{d}$ with an a-priori estimate,  $\tilde{\mu}_{d}$, the ``anticipated effect size'', and replace $\sigma^{2}$ with an a-priori estimate,  $\tilde{\sigma}^{2}$, the ``anticipated variance''. Then one might be interested in calculating six values: the probabilities of obtaining each of the three possible results (positive, negative and inconclusive) under two hypothetical scenarios, (1) where ${\mu}_{d}=0$, and (2) where ${\mu}_{d}$ equal to $\tilde{\mu}_{d}$, the value expected given results in the literature.  One might also be interested in a hybrid approach whereby one specifies a composite null and alternative distribution.  Since the objective of any study should be to obtain a conclusive result, sample size could also be calculated with the objective to maximize the likelihood of success, i.e. to minimize $Pr(inconclusive)$.

\begin{figure}[h!]
\begin{centering}
\fbox{\includegraphics[width=13cm]{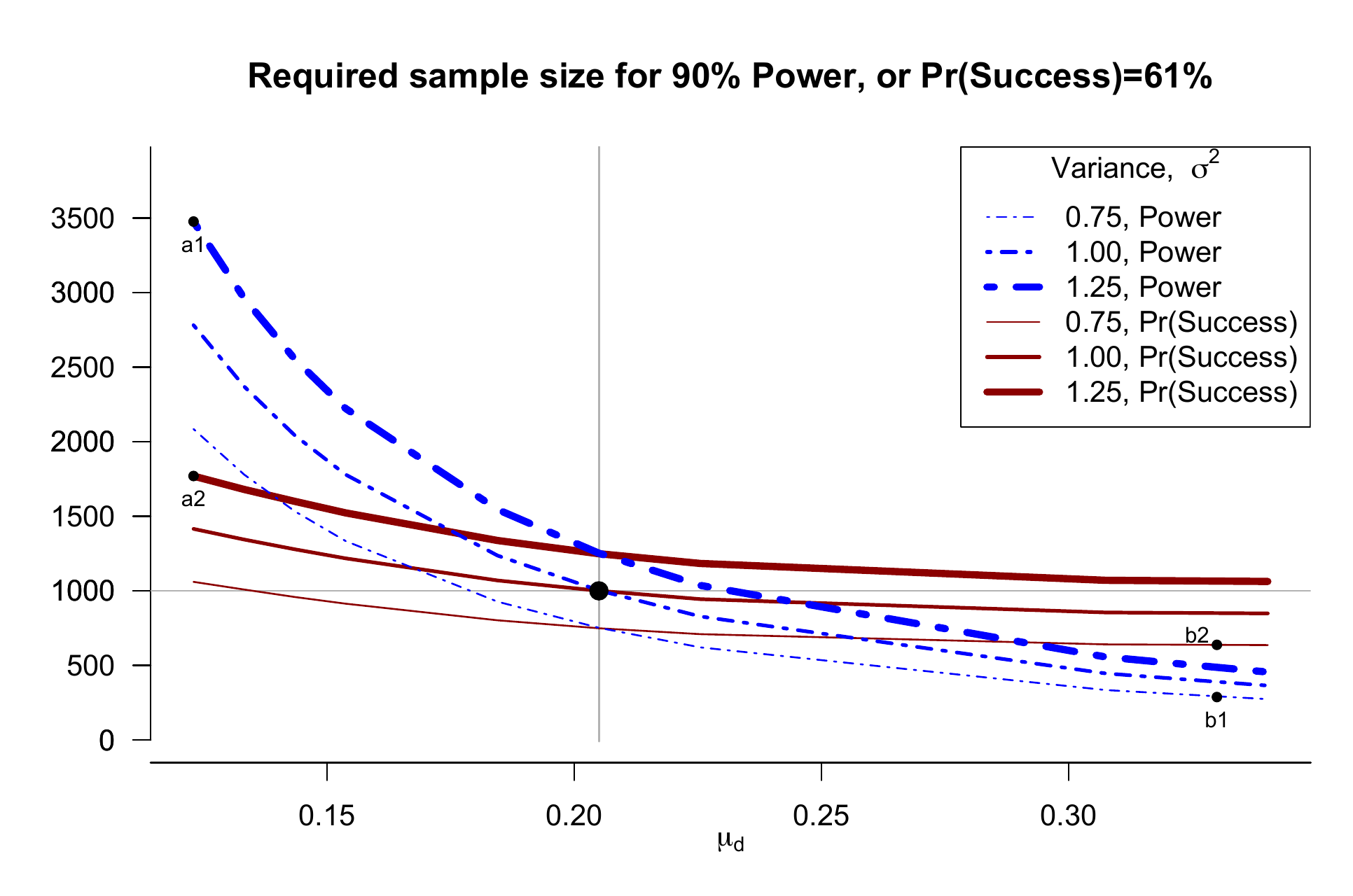}}
\caption{\footnotesize{Suppose the anticipated effect size is $\tilde{\mu}_{d}=0.205$, the anticipated variance is  $\tilde{\sigma}^{2}= 1$ and the equivalence margin is pre-specified with $\Delta=0.1025$ (=$\frac{1}{2}\tilde{\mu}_{d}$).   Then, based on the desire for Pr(positive) = 90\% (i.e. ``power'' = 0.90) (dashed blue lines) or a $61\%$ ``probability of success'' (solid red lines) a sample size of $n=1,000$ (with $n_{1}=n_{2}$) would be required. If the true variance is slightly larger than anticipated, ${\sigma}^{2}= 1.25$, and the effect size smaller, $\mu_{d}=0.123$, the actual sample size needed for 90\% power is in fact 3,476, while the actual sample size needed for $Pr(success) = 61\%$ is 1,770; see points ``a1'' and ``a2''.  On the other hand, if the true variance is slightly smaller than anticipated, ${\sigma}^{2}= 0.75$ and the effect size greater, $\mu_{d}=0.33$, the actual sample size needed for 90\% power is only 288, while the actual sample size needed for $Pr(success | \delta, d, s) = 61\%$ is 638; see points ``b1'' and ``b2''.}}
\label{fig:samplesize}
\end{centering}
\end{figure}

 \noindent The quantity:
 \begin{equation}
Pr(success) =1 - \frac{1}{2}Pr(inconclusive ; {\mu}_{d}=\tilde{\mu}_{d}, \tilde{\sigma}^{2}) - \frac{1}{2}Pr(inconclusive ; {\mu}_{d}=0, \tilde{\sigma}^{2})
\end{equation}

 \noindent represents the probability of a ``successful'' study under the assumption that the null ($\mu_{d}=0$)  and alternative (with specified $\mu_{d}=\tilde{\mu}_{d}$) are equally likely.  (Note that both false positive-, and false negative- studies in this equation are considered ``successful''.)  This weighted average could be considered a simple version of what is known as ``assurance'' \citep{o2005assurance}. Figure \ref{fig:samplesize} shows how consideration of $Pr(success)$ as the criteria for determining sample size attenuates the effect of $\tilde{\mu}_{d}$ on the required sample size.  Suppose one calculates that the required sample size is $n=1,000$ based on the desire for 90\% statistical power and the belief that  $\tilde{\sigma}^{2}$= 1 and $\mu_{d}$=0.205, with $n_{1}=n_{2}$ as before.  This corresponds to a $61\%$ probability of success for $\Delta$=0.1025 (=$\frac{1}{2}\mu_{d}$).  If the true variance is slightly larger than anticipated, ${\sigma}^{2}$=1.25, and the difference in means smaller, $\mu_{d}$=0.123, the actual sample size needed for 90\% power is in fact $n$=3,476, while the actual sample size needed for $Pr(success | \Delta, d, s) = 61\%$ is $n$=1,770.  On the other hand, if the true variance is slightly smaller than anticipated, ${\sigma}^{2}= 0.75$ and the difference in means greater, $\mu_{d}$=0.33, the actual sample size needed for 90\% power is only $n$=288, while the actual sample size needed for $Pr(success | \delta, d, s)$=61\% is $n$=638.  It follows that, if one has little certainty in $\mu_d$ and $\sigma^{2}$, calculating the required sample size with consideration of $Pr(success)$ may be less risky.

For related work on statistical power, see \citet{shao2008hybrid} who propose a hybrid Bayesian-frequentist approach to evaluate power for testing both superiority and non-inferiority.  \citet{jia2015sample} discuss a related sample size planning approach that considers both statistical significance and clinical significance.  Finally, \citet{jiroutek2003new} advocate that, rather than calculate statistical power (the probability of rejecting $\theta=\theta_{0}$ should the alternative be true), one should calculate the probability that the width of a confidence interval is less than a fixed constant and the null hypothesis is rejected, given that the confidence interval contains the true parameter.

\section{A comparison with Bayesian testing}

Recently, Bayesian statistics have been advocated for, as a ``possible solution to publication bias'' \citep{konijn2015possible}.  In particular, there have been many Bayesian testing schemes proposed in the psychology literature; see the discussion of \citet{mulder2016editors} and, for an accessible overview of the ``Bayesian $t$-test'', see \citet{gonen2010bayesian}.  What's more, publication policies based on Bayesian testing schemes are currently in use by a small number of journals and are the preferred approach for some (e.g. \citet{dienes2017four}).   In response to these developments, we will compare, with regards to their operating characteristics, CET and a Bayesian testing scheme. This brings to mind \citet{dienes2014using} who compares testing with Bayes Factors (BF) to testing with ``interval methods'' and notes that with interval methods, a study result is a ``reflection of the data'', whereas with BFs the result reflects the ``evidence of one theory over another''.   What follows is a brief overview of one Bayesian scheme and an investigation of how it compares to CET. 

The Bayes Factor is a valuable tool for determining the degree of evidence for the absence of a treatment effect, see most recently \citet{hoekstra2017bayesian}.  Consider, for the two-sample testing of normally distributed data (as described for box 2.1), a Bayes Factor testing scheme in which we take the JZS (Jeffreys-Zellner-Siow) prior for the alternative hypothesis, see \citet{rouder2009bayesian}.  Note that, for the Bayes Factor, the null hypothesis, $H_{0}$, corresponds to $\mu_{d}=0$; and the alternative, $H_{1}$, corresponds to $\mu_{d} \ne 0$.

The JZS testing scheme involves placing a normal prior on $\eta=(\mu_2 - \mu_1)/\sigma$, $\quad  \eta \sim Normal(0, \sigma^{2}_{\eta})$, and for the hyper-parameter $\sigma_{\eta}$, placing an inverse chi-squared prior, $\sigma^{2}_{\eta} \sim inv.\chi^{2}(1)$.  Integrating out $\sigma_{\eta}$ shows that this is equivalent to having a Cauchy prior, $\eta \sim Cauchy$.   The JZS prior is recommended as a reasonable ``objective prior'' to be used in a Bayesian alternative to the common frequentist $t$-test \citep{rouder2009bayesian}.  We can write the JZS Bayes Factor in terms of the standard $t$-statistic, $T = (\bar{x}_{1}-\bar{x}_{2})/s^{*}$, with $n^{*} = n_{1}n_{2}/(n_{1}+n_{2})$, as follows:


\begin{equation}
B_{01} = \frac{(1 + \frac{T^{2}}{n^{*}-1})^{-(n^{*}/2)}}{\int_{0}^{\infty}(1+n^{*}g)^{-1/2} \Big(1+ \frac{T^{2}}{(1+n^{*}g)(n^{*}-1)} \Big)^{-(n^{*}/2)} (2\pi)^{-1/2} g^{-3/2} e^{-1/(2g)} dg }
\end{equation}

\begin{figure}[h!]
\begin{centering}
\fbox{\includegraphics[width=16cm]{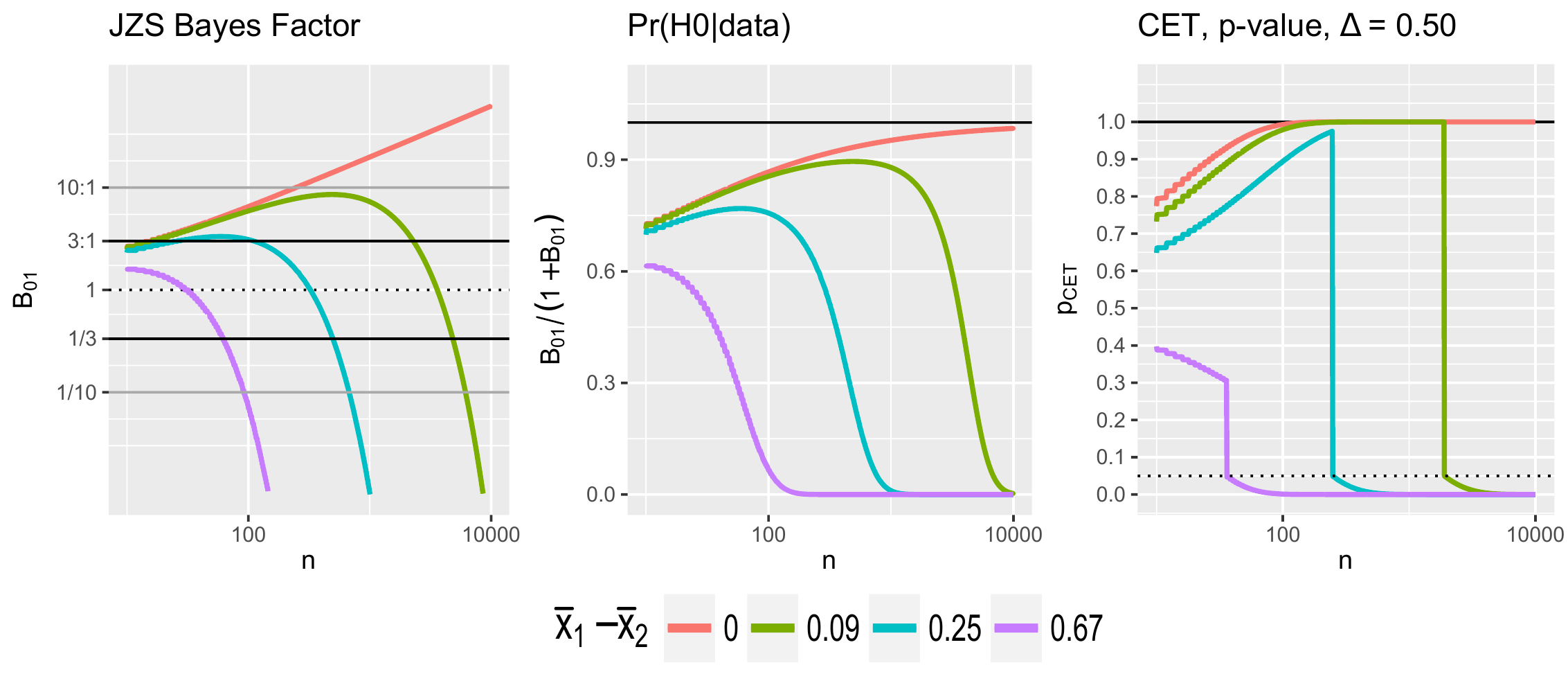}}
\caption{\footnotesize{Based on Figure 5 from  \citep{rouder2009bayesian}.  Left (middle; right) panel shows how the JZS Bayes Factor (the posterior probability of $H_{0}$; $p_{CET}$) changes with sample size for four different observed mean differences, $\hat{\mu}_{d}=\bar{x}_{1}-\bar{x}_{2}$.  The observed variance is constant, $s^{2}_{p}=1$. }}
\label{fig:B_CET}
\end{centering}
\end{figure}

Figure \ref{fig:B_CET} -left panel shows how the JZS Bayes Factor changes with sample size for four different values of the observed difference in means, $\hat{\mu}_{d}=\bar{x}_{1} -\bar{x}_{2}$; the observed variance remains constant at $s^{2}_{p}=1$.  When the observed mean difference is exactly 0, the BF increases logarithmically with $n$.   For small to moderate $\hat{\mu}_{d}$, the BF supports the null for small values of $n$ but, as $n$ becomes larger, yields less support for the null and eventually favours the alternative.   The horizontal lines mark the 3:1 and 1/3 thresholds (``moderate evidence'') as well as the 10:1 and 1/10 thresholds (``strong evidence'').  

Figure \ref{fig:B_CET} -right panel shows  $p_{CET}$-values for the same four values of $\hat{\mu}_{d}$, constant $s^{2}_{p}=1$, and the equivalence margin of $[-0.50,0.50]$. The lines suggest that CET possesses similar ``ideal behaviour'' \citep{rouder2009bayesian} as is observed with the BF.   When the observed mean difference is exactly zero, CET provides increasing evidence in favour of equivalence with increasing $n$.  For small to moderate $\hat{\mu}_{d}$, CET supports the null at first and then, as $n$ becomes larger, at a certain point favours the alternative.  The sharp change-point represents border cases.  Consider case ``f''  in Figure 1: if $n$ increased, the confidence intervals would shrink and at a certain point, the $(1-\alpha_{1})\%$ C.I. would be exclude 0 (similar to case ``b'').  At that point, the result abruptly changes from ``negative'' to ``positive''.  While this abrupt change may appear odd at first, it may in fact be more desirable than the smooth transition of the BF.  Consider for example when $\hat{\mu}_{d}=0.25$.  Then for $n$ between 112 and 496, the BF will be strictly above 1/3 and strictly below 3, and as such the result, by BF, is inconclusive.  In contrast, for the same range in sample size, $p_{CET}$ will be either above 0.90 or below 0.05.  As such, with $\alpha_{1}=0.05$ and $\alpha_{2}=0.10$, a conclusive result is obtained.  While careful interpretation is required (e.g. ``the effect is significant yet not of a meaningful magnitude''), this may be preferable in some settings to the BF's inconclusive result.

We can also consider the posterior probability of $H_{0}$ (i.e. $\mu_{d}=0$) equal to $B_{01}/(1+B_{01})$ (when the prior probabilities $Pr(H_{0})$ and $Pr(H_{1})$ are equal), plotted in Figure \ref{fig:B_CET} -middle panel.   The similarities and differences between $p$-values and posterior probabilities have been widely discussed; see \citet{berger1987testing} and more recently \citet{greenland2013living} and \citet{marsman2017three}.  Figure 5 suggests that the JZS-BF and CET testing may often result in similar conclusions.  We investigate this further by means of a simple simulation study.
 

\subsection{Simulation Study}


We conducted a small simulation study to compare the operating characteristics of testing with the JZS-BF relative to with the CET approach.   CET conclusions were based on setting $\Delta=0.50$, $\alpha_{1}$=0.05 and $\alpha_{2}$=0.10.  JZS BF conclusions were based on a threshold of 3 or greater for evidence in favour of the a negative result and less than 1/3 for evidence in favour of a positive result.  BFs in the 1/3 - 3  range correspond to an inconclusive result.  A threshold of 3:1 can be considered ``substantial evidence'' \citep{wagenmakers2011psychologists}.  Note that one advantage of Bayesian methods, is that sample sizes need not be determined in advance; see \citet{rouder2014optional}.    \citet{schonbrodt2016bayes} list three ways one might design sample size for a study using the BF for testing.  For the simulation study here we examine only the ``fixed-$n$ design''.

\begin{figure}[h!]
\begin{centering}
\fbox{\includegraphics[width=16cm]{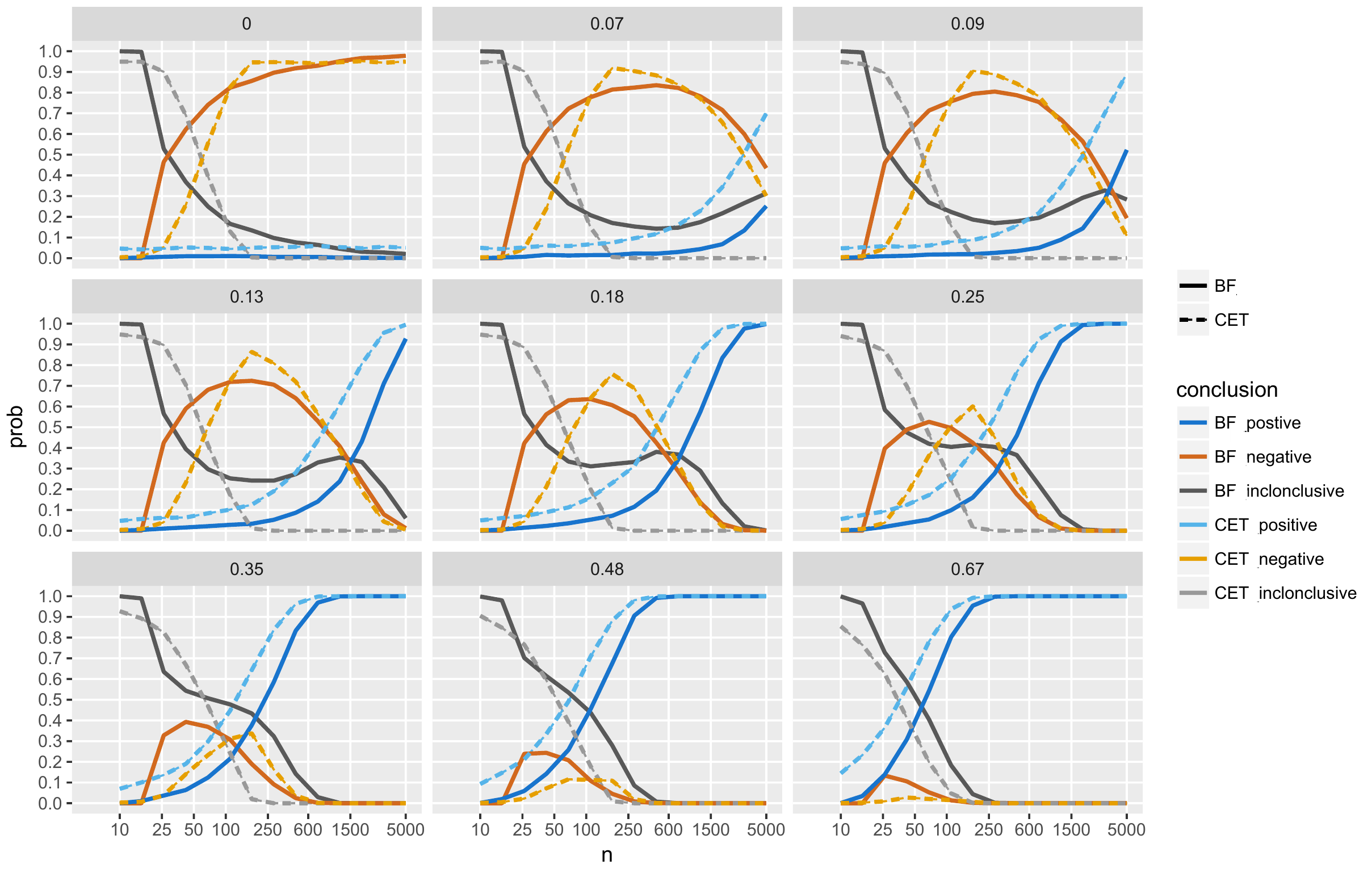}}
\caption{\footnotesize{The probability of obtaining each conclusion by Bayesian testing scheme (JZS-BF with fixed sample size design,  BF threshold of 3:1) and CET ($\alpha_{1}=0.05, \alpha_{2}=0.10$).  Each panel displays the results of simulations with true mean difference, $\mu_{d}=$ 0, 0.07, 0.09, 0.13, 0.18, 0.25, 0.35, 0.48, and 0.67. }}
\label{fig:simstudy1}
\end{centering}
\end{figure}

 For a range of $\mu_{d}$ (= 0, 0.07, 0.09, 0.13, 0.18, 0.25, 0.35, 0.48, 0.67) and 14 different sample sizes ($n$ ranging from 10 to 5,000, with $n_{1}=n_{2}$) we simulated normally distributed two-sample datasets (with $\sigma^{2}=1$).  For each dataset, we obtained CET $p$-values, JZS BFs and declared the result to be positive, negative or inconclusive accordingly.   Results are presented in Figure \ref{fig:simstudy1}, based on 5,000 distinct simulated datasets per scenario.   
 
 Several  findings merit comment:


 \begin{itemize}
 \item{ In this simulation study, the JZS-BF admits a very low frequentist type I error, recorded at most $\approx 0.01$, for a sample size of $n=110$.  As the sample size increases, the frequentist type I error diminishes to a negligible level. }
 \item{The JZS-BF requires less data to reach a negative conclusion than the CET.  However, with moderate to large sample sizes ($n$=100 to 5,000) and small true mean differences ($\mu_{d}=$ 0 to 0.25), both methods are approximately equally likely to deliver a negative conclusion.}
  \item{While the JZS-BF requires less data to reach a conclusion when the true mean difference is small ($\mu_{d}=$ 0 to 0.25) (see how solid black curve drops more rapidly than the dashed grey line), there are scenarios in which larger sample sizes will surprisingly reduce the likelihood of obtaining a conclusive result (see how the solid black curve drops abruptly then rises slightly as $n$ increases for $\mu_{d}=$ 0.07, 0.09, 0.13, and 0.18.) }
   \item{The JZS-BF is always less likely to deliver a positive conclusion (see how dashed blue line is always higher than solid blue line).  In the scenarios like those considered, JZS-BF may require larger sample sizes for reaching a positive conclusion and may be considered ``less powerful'' in a traditional frequentist sense.}
   \end{itemize}


The results of the simulation study suggest that, in many ways, the JZS-BF and CET operate very similarly.  Think of JZS-BF and CET as two pragmatically similar, yet philosophically different, tools for making ``trichotomous significance-testing decisions''.  Both tools will often result in the same outcome, given the same data.

\section{A CET publication policy}

Many researchers have put forth ideas for new publication policies aimed at addressing the issue of publication bias.  There is a wide range of opinions on how to incorporate more null results into the published literature.  Consider just a few interesting ideas.  In an editorial titled ``Journals Should Publish All \emph{Null} Results and Should Sparingly Publish \emph{Positive} Results.'', \citet{ioannidis2006journals} writes: ``\emph{Null} results should be published promptly in print in short versions, with more extensive details in web-based files. \emph{Positive} results should be published equally promptly, but only on the web, pending independent replication; once refuted, the original article and the refutation could be printed as a single nice null report; the rare validated findings should appear in print with full details.''  Another suggestion is that of \citet{shields2000publication} who advocates accepting null papers in a special section of a journal, under the category of ``Null Results in Brief''.  The null papers in this section, would provide only a brief summary of the methods and results of the studies. With regards to power, \citet{shields2000publication} states that: ``For the [null] paper to be considered for publication, there must be sufficient statistical power to test the a-priori hypothesis. For example, the authors should state the level of power to detect an odds ratio of 2.0 with the current sample size.''   \citet{dirnagl2010fighting} makes a similar suggestion with the note that ``the quality of the data submitted to our Negative Results section must meet the same rigorous standards that our journal applies to all other submissions. In fact, it may be said that the standards must even exceed those applied currently, as type II error (false negatives) considerations need to be included.''

 \begin{figure}[h!]
\begin{centering}
\fbox{\includegraphics[width=10cm]{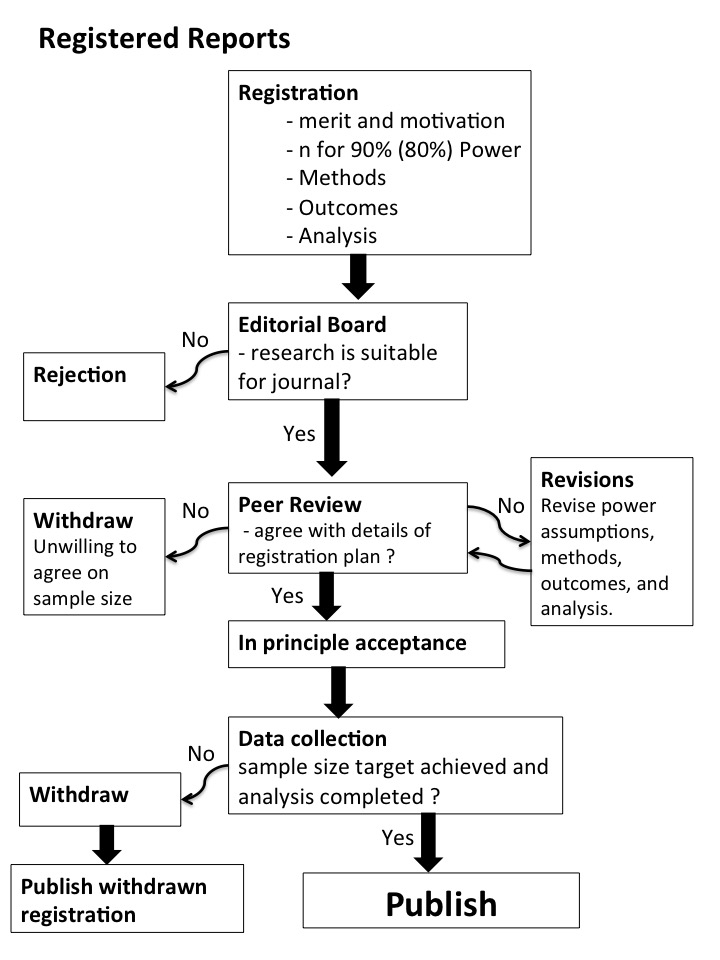}}
\caption{\footnotesize{The RR publication policy.}}
\label{fig:RR}
\end{centering}
\end{figure}

Certainly one of the most exciting proposals of late is that of \emph{Registered Reports} (RR).  RR is one of many proposed ``two-step''  manuscript review schemes in which acceptance for publication is granted prior to obtaining the results, see e.g. \citet{walster1970proposal}, \citet{lawlor2007quality} as well as more recently \citet{mell2014introducing} and  \citet{smulders2013two}. Figure  \ref{fig:RR} illustrates the RR procedure and \citet{chambers2014instead} provide an in-depth explanation answering a number of frequently asked questions.    The RR policy has two central components: (1) pre-registration and (2) the ``RR commitment to publish''.  

Pre-registration can be extremely beneficial as it reduces ``researcher degrees of freedom'' \citep{simmons2011false} and prevents, to a large degree, many questionable research practices including data-dredging \citep{berry1990subgroup}, the post-hoc fabrication of hypotheses, (``HARKing'') \citep{kerr1998harking}, and $p$-hacking \citep{gelman2013garden}.  However, on its own, pre-registration does little to prevent publication bias.  This is simply because pre-registration: (a1) cannot prevent authors from disregarding negative results \citep{song2014medical}; (a2) does nothing to prevent reviewers and editors from rejecting studies for lack of significance; and (a3) does not guarantee that peer reviewers  consider compliance with the pre-registered analysis plan \citep{van2015differences, mathieu2013use, chan2008discrepancies}.\footnote{This paragraph based in part on the post from ``ff524'' at https://academia.stackexchange.com/questions/74711/why-isnt-pre-registration-required-for-all-experiments.} 

Consider the field of medicine as a case study.   For over a decade, pre-registration of clinical trials has been required by major journals as a prerequisite for publication.  Despite this heralded policy change, selective outcome reporting remains ever prevalent, \citep{ramsey2008commentary, mathieu2009comparison, ross2009trial, huic2011completeness}. (This being said, new 2017/2018 guidelines for the \emph{clinicaltrial.gov} registry show much promise in addressing a1, a2, and a3; see \citet{zarin2016trial}.)


In order to prevent publication bias, RR complements pre-registration with a ``commitment to publish''.   In practice this consists of an ``in principle acceptance'' policy along with the policy of publishing ``withdrawn registration" (WR) studies.  In order to counter authors who may simply shelve negative results following pre-registration (a1), RR journals commit to publishing the abstracts of all withdrawn studies as WR papers.  By guaranteeing that, should a study follow its pre-registered protocol, it will be accepted for publication (``in principle acceptance''), RR prevents reviewers and editors from rejecting a study based on the results (a2).  Finally, RR requires that a study is in strict compliance with the pre-registered protocol if it is to be published (a3).   In order to keep a RR journal relevant (not simply full of inconclusive studies), RR requires, as part of registration, a researcher commit to a sample size large enough of achieve 90\% (in some cases 80\%) statistical power.  In a small number of RR journals, a Bayesian alternative option is offered.  Instead of committing to a specific sample size, researchers commit to achieving a certain BF. 

The policy we put forth here is not meant to be an alternative to the ``pre-registration'' component of RR.  Its benefits are clear, and in our view is most often ``worth the effort'' \citep{wager2013hardly}.  If implemented properly, pre-registration should not ``stifle exploratory work'' \citep{gelman2013preregistration}.  Instead, what follows is an alternative to the second ``commitment to publish'' component.

\subsection{Outline of a CET-based publication policy}

 \begin{figure}[h!]
\begin{centering}
\fbox{\includegraphics[width=11cm]{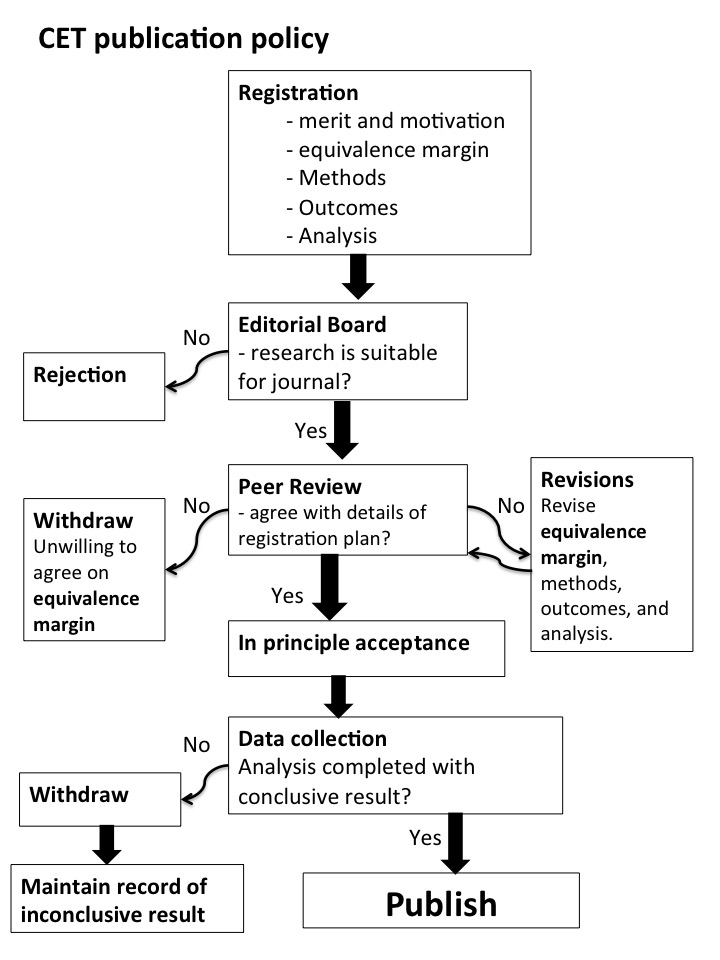}}
\caption{\footnotesize{The CET publication policy.}}
\label{fig:RR2}
\end{centering}
\end{figure}

Figure \ref{fig:RR2} illustrates the steps of our proposed policy.  What follows is a general outline.

\textbf{Registration-}
In the first stage of a CET-based policy, before any data are collected, a researcher will register the intent to conduct a study with a journal's editor.  As in a RR policy, this registration process will detail the motivations and merits of the study and list the defined outcomes, the various hypotheses to be tested, and the proposed methods for analysis.  Unlike in the RR policy, the registration will not require a sample size calculation showing a specific level of statistical power.  Instead, the researcher will define an equivalence margin for each hypothesis test to be carried out.  For example if a researcher intends to fit a linear regression model with five explanatory variables, a margin should be defined for each of the five variables.   A target sample size should be stated, but need not be justified with regards to power considerations.  The researcher will also need to note if there are plans for any sample size reassessments, interim and/or futility analyses.

\textbf{Editorial and Peer Review-} If the merits of the study satisfy the editorial board, the registration study plan will then be sent to reviewers to assess whether the methods for analysis are adequate and whether the equivalence margins are sufficiently narrow.   Once the peer reviewers are satisfied (possibly after revisions to the registration plan), the journal will then agree, in principle, to accept the study for eventual publication, on condition that either a positive or a negative result is obtained.

\textbf{Data collection-} Armed with this ``in principle acceptance'',  the researcher will then collect data in an effort to meet the established sample size target.  Once the data are collected and analyses complete, the study will be published if and only if either a positive or negative result is obtained as defined by the pre-specified equivalence margins.  Inconclusive results will not generally be published thus protecting a journal from becoming a ``dumping ground'' for failed studies.  In very rare circumstances, however, it may be determined that an inconclusive study offers a substantial contribution to the field and should therefore be considered for publication.  
  As is required practice for good reporting, any failure to meet the target sample size should be stated clearly, along with a discussion of the reasons for failure and the consequences with regards to the interpretation of results; see \citet{toerien2009review}.

This proposed policy is most similar to the RR Bayesian option outlined in \mbox{Section 3}.  Journals only commit (``in principle acceptance'') to publish conclusive studies (small $p$-value/ BF above or below a certain threshold) and no a-priori sample size requirements are forced upon a researcher.  As noted in Section 3, we expect that the conclusions obtained under either policy will often  be the same.  The CET-based policy therefore represents a frequentist alternative to the RR-BF policy.   

A journal may wish to require stricter or weaker requirements for publication of results and this can be done by setting thresholds for $\alpha_{1}$ and $ \alpha_{2}$ accordingly, (e.g. stricter thresholds: $\alpha_{1}=0.01, \alpha_{2}=0.05, \Delta=0.5s_{p}$; weaker thresholds: $\alpha_{1}=0.05, \alpha_{2}=0.10, \Delta=0.5s_{p}$).  Also, note that a $p$-value may be used in one way for the interpretation of results and another way to inform publication decisions.  Recently, a large number of researchers \citep{benjamin2017redefine} have publically suggested that the threshold for defining ``statistical significance'' be changed from $p<0.05$ to $p<0.005$.  However, they are careful to emphasize that while this stricter threshold should change the description and interpretation of results, it should not change ``standards for policy action nor standards for publication''.

\textbf{``Gaming the system''-} When evaluating a new publication policy one should always ask how (and how easily) a researcher could -unintentionally or not- ``game the system''.  For the CET-policy as described above, we see one rather obvious strategy.  

Consider the following, admittedly extreme, example.  A researcher submits 20 different study protocols for pre-registration each with very small sample size targets (e.g. $n=8$).  Suspend any disbelief, and suppose these studies all have good merit, are all well written, and are all well designed (besides being severely underpowered), and are therefore all granted in principle acceptance.  Then, in the event that the null is true (i.e. $\mu=0$) for all 20 studies (and with $\alpha_{1}=0.05$), it is expected that at least one study out of the 20 will obtain a positive result and thus be published.  This is publication bias at its worst.

In order to discourage this unfortunate practice, we suggest making a researcher's history of pre-registered studies available to the public.  The researcher ``gaming the system'' in this way will still score his one ``type I error'' publication, but it will also be known to the public that in 19 other experiments, research was unsuccessful and valuable resources were essentially wasted.   With digital identifiers such as ORCID \citep{haak2012orcid}, it should be straightforward to maintain a track record across journals and disciplines of the number of successful and unsuccessful studies for each researcher/laboratory.  However, while potentially beneficial, we do not anticipate this type of action being necessary.  With the CET-policy, it is no longer in a researcher's interest to ``game the system''.  

Consider once again, our extreme example.  The strategy has an approximately 64\% chance of obtaining at least one publication at a cost of 20 submissions and a total of $160=8\cdot20$ observations.  If the goal is to maximize the probability of being published \citep{charlton2006should}, then it is far more efficient to submit a single study with $n=160$, in which case there is an approximately 98\% chance of obtaining a publication (i.e. with $\alpha_{2}=0.10$, $Pr(inconclusive|\mu=0, \sigma^{2}=1, \Delta=0.5)=0.02$; 92\% chance with $\alpha_{2}=0.05$,  $Pr(inconclusive|\mu=0, \sigma^{2}=1, \Delta=0.5)=0.08$).

\section{Conclusion}

Publication bias has been recognized as a serious problem for several decades now \citep{rosenthal1979file}.  Yet, in many fields, it is only getting worse \citep{pautasso2010worsening}.  An investigation by \citet{kuhberger2014publication} concludes that the ``entire field of psychology'' is now tainted by ``pervasive publication bias''.  

 There remains substantial disagreement on the merits of pre-registration and result-blind peer-review (see e.g. \citet{coffman2015pre, de2013selective, van2014publishing}).  Yet, all can agree that innovative publication policy prescriptions can be part of the solution to the ``reproducibility crisis''.  While some call for dropping $p$-values and strict thresholds of evidence altogether, we believe that it is not worthwhile to fight ``the temptation to discretize continuous evidence and to declare victory'' \citep{gelman2017some}. Instead, the research community should embrace this ``temptation'' and work with it to achieve desirable outcomes.  Indeed, one way to address the ``practical difficulties that reviewers face with null results'' \citep{findley2016can} is to \emph{further} discretize continuous evidence by means of equivalence testing and we submit that CET can be an effective tool for distinguishing those ``high-quality null results'' \citep{shields2009null} worthwhile of publication.

Recently, a number of influential researchers \citep{mcshane2017abandon} have argued that to address low reliability, scientists, reviewers and regulators should ``abandon statistical significance''.  (Somewhat ironically, in some fields, such as reinforcement learning, the currently proposed solution is just the opposite: the adoption of ``significance metrics and tighter standardization of experimental reporting'' \citep{henderson2017deep}.)  We recognize that current publication policies, in which evidence is dichotomized (without any ``ontological basis'') may be highly unsatisfactory.  However, one benefit to adopting distinct categories based on clearly defined thresholds (as in the CET-policy) is that one can assess, in a systematic way, the state of published research (in terms of reliability, power, reproducibility, etc.).  While using a ``more holistic view of the evidence'' to inform publication decisions may (or may not?) prove effective, `meta-research' under such a paradigm is clearly less feasible.  As such, the question of effectiveness may perhaps never be adequately answered.

Bayesian approaches offer many benefits.  However, we see three main drawbacks.  First, adopting Bayesian testing requires a substantial paradigm shift.  Since the interpretation of findings deemed significant by traditional NHST may differ with Bayes, some will no doubt be reluctant to accept the shift.  With CET, the traditional usage and interpretation of the $p$-value remains unchanged, except in circumstances when one fails to reject the null.  As such, CET does not change the interpretation of findings already established as significant.    Indeed, CET simply ``extend[s] the arsenal of confirmatory methods rooted in the frequentist paradigm of inference'' \citep{wellek2017critical}.  Second, as we observed in our simulation study, Bayesian testing is potentially less powerful than NHST (in the traditional frequentist sense, with a fixed sample size design) and as such could require substantially larger sample sizes.  Finally, we share the concern of \citet{morey2011bayes} who write that Bayesian testing ``provides no means of assessing whether rejections of the nil [null] are due to trivial or unimportant effect sizes or are due to more substantial effect sizes.''  For these reasons, we believe CET should be welcomed by any ``pragmatic Bayesian'' \citep{kass2006kinds}. 

There are many potential areas for further research.  Determining whether $\Delta$ and/or $\alpha_1$ and/or $\alpha_2$ should be chosen with consideration of the sample size is important and not trivial; related work includes \citet{perez2014changing} who put forward a ``Bayes/non-Bayes compromise'' in which the $\alpha$-level of a confidence interval changes with $n$.  Issues which have proven problematic for standard equivalence testing must also be addressed for CET.  These include multiplicity control \citep{lauzon2009easy} and potential problems with interpretation \citep{abereggempirical}.  It would also be worthwhile considering whether CET is appropriate for testing for baseline balance, \citet{senn1994testing}. Finally, the impact of a CET policy on meta-analysis should be examined, \citet{hedges1992modeling} (i.e. how should one account for the exclusion of inconclusive results in the published literature when deriving estimates in a meta-analysis?).

The publication policy outlined here should be welcomed by journal editors, researchers and all those who wish to see more reliable science.  Research journals which wish to remain relevant and gain a high impact factor should welcome the CET-policy as it offers a mechanism for excluding inconclusive results while providing a space for potentially impactful negative studies.  Embracing equivalence testing is an effective way to make publishing null results ``more attractive'' \citep{o2011negative}.  Researchers should be pleased with a policy that provides ``in principle acceptance'' and does not insist on specific sample size requirements that may not be feasible or desirable.  

The requirement to specify an equivalence margin prior to collecting data will have the additional benefit of forcing researchers and reviewers to think about what would represent a meaningful effect size before embarking on a given study.  While there will no doubt be pressure on researchers to ``$p$-hack'' in order to meet either the the $\alpha_{1}$ or $\alpha_{2}$ threshold, this can be discouraged by insisting that an analysis strictly follows the pre-registered analysis plan.  Adopting strict thresholds for significance can also act as a deterrent.  Finally, we believe that the CET-policy will improve the reliability of published science by not only allowing for more negative research to be published, but by modifying the incentive structure driving research \citep{nosek2012scientific}.

Using an optimality model, \citet{higginson2016current} conclude that, given current incentives, the rational strategy of a scientist is to ``focus almost all of their research effort on underpowered exploratory work [... and] carry out lots of underpowered small studies to maximize their number of publications, even though this means around half will be false positives.''  This result is in line with the views of many (e.g. \citet{bakker2012rules}, \citet{button2013power} and \citet{gervais2015powerful}), and provides the basis for why statistical power in many fields has not improved \citep{smaldino2016natural} despite being highlighted as an issue over six decades ago \citep{cohen1962statistical}.  A CET-based policy may provide the incentive scientists need to pursue higher statistical power.  If CET can change the incentives driving research, the reliability of science will be further improved.   More research on this question (i.e. ``meta-research'') is needed.




\acks{We gratefully acknowledge support from Natural Sciences and Engineering Research Council of Canada.  We also wish to thank Drs. John Petkau and Will Welsh for their valuable feedback.}


\bibliographystyle{abbrv}
\bibliography{truthinscience}

\end{document}